\documentclass[sigconf]{acmart}
\renewcommand\footnotetextcopyrightpermission[1]{} 
\AtBeginDocument{%
  }

\setcopyright{acmlicensed}
\copyrightyear{2018}
\acmYear{2018}
\acmDOI{XXXXXXX.XXXXXXX}
\acmConference[Conference acronym 'XX]{Make sure to enter the correct
  conference title from your rights confirmation email}{June 03--05,
  2018}{Woodstock, NY}
\acmISBN{978-1-4503-XXXX-X/2018/06}

\usepackage{graphicx}

\usepackage{subcaption}

\usepackage{amsfonts}

\usepackage{booktabs} 

\usepackage{stmaryrd}

\usepackage{multirow}

\usepackage{makecell}

\usepackage{amsmath}

\usepackage{algorithm}
\usepackage{algorithmic}

\begin{document}

\title{Hermes: A Unified High-Performance NTT Architecture with Hybrid Dataflow}

\author{Hang Gu}
\affiliation{%
  \institution{School of Software Engineering, University of Science and Technology of China}
  \city{Hefei}
  \country{China}
}

\author{Teng Wang}
\authornote{Corresponding author.}
\affiliation{%
  \institution{Suzhou Institute for Advanced Research, University of Science and Technology of China}
  \city{Suzhou}
  \country{China}
}

\author{Qianyu Cheng}
\affiliation{%
  \institution{School of Computer Science and Technology, University of Science and Technology of China}
  \city{Hefei}
  \country{China}
}

\author{Jinao Li}
\affiliation{%
  \institution{School of Computer Science and Technology, University of Science and Technology of China}
  \city{Hefei}
  \country{China}
}

\author{Zhendong Zheng}
\affiliation{%
  \institution{School of Computer Science and Technology, University of Science and Technology of China}
  \city{Hefei}
  \country{China}
}

\author{Lei Gong}
\affiliation{%
  \institution{School of Computer Science and Technology, University of Science and Technology of China}
  \city{Hefei}
  \country{China}
}

\author{Wenqi Lou}
\affiliation{%
  \institution{Suzhou Institute for Advanced Research, University of Science and Technology of China}
  \city{Suzhou}
  \country{China}
}

\author{Xi Li}
\affiliation{%
  \institution{School of Computer Science and Technology, University of Science and Technology of China}
  \city{Hefei}
  \country{China}
}

\author{Xuehai Zhou}
\affiliation{%
  \institution{School of Computer Science and Technology, University of Science and Technology of China}
  \city{Hefei}
  \country{China}
}
\affiliation{%
  \institution{Suzhou Institute for Advanced Research, University of Science and Technology of China}
  \city{Suzhou}
  \country{China}
}

\renewcommand{\shortauthors}{Gu et al.}

\begin{abstract}

Fully Homomorphic Encryption (FHE) relies heavily on the Number Theoretic Transform (NTT), making NTT a major performance bottleneck due to its intensive polynomial computations. Hybrid Homomorphic Encryption (HHE), which integrates arithmetic and logic FHE, further requires support for multiple NTT lengths. However, existing accelerators mainly optimize NTT throughput and do not provide unified support for HHE. This paper presents Hermes, a unified high-performance NTT architecture based on hybrid dataflow. Hermes exploits parallelism along both temporal and spatial dimensions and incorporates a fully pipelined on-chip computing core. A conflict-free on-chip fragmentation algorithm is introduced to resolve bank conflicts and enable burst HBM access, while an efficient dataflow improves computational intensity through data reuse, reducing bandwidth demand. Experimental results show that Hermes supports multiple NTT lengths and achieves up to 13.6× and 1.3× higher throughput than state-of-the-art GPU and FPGA accelerators, respectively. Our source code is available at \textbf{\url{https://anonymous.4open.science/r/Hermes_conf-4E6F}}.
\end{abstract}


  


\maketitle

\section{Introduction}
The Number Theoretic Transform (NTT), a finite-field variant of the Discrete Fourier Transform (DFT)~\cite{James1965CT,Gentleman1966GS}, is a key algorithm for accelerating polynomial multiplication in modern cryptography. By exploiting modular $n$-th roots of unity, the NTT converts polynomial multiplication into point-wise multiplication followed by an inverse transform, reducing the complexity from $\mathcal{O}(n^{2})$ to $\mathcal{O}(n\log n)$. With this quasi-linear efficiency, the NTT has become a fundamental primitive in lattice-based cryptography (e.g., ring-LWE~\cite{lyubashevsky2010ideal}) and is widely used for fast convolution over large integers and polynomials. In many post-quantum cryptographic schemes, polynomial multiplication is a dominant performance bottleneck, and NTT-based acceleration significantly improves overall efficiency. Therefore, in advanced applications such as Fully Homomorphic Encryption (FHE)~\cite{gentry2009fully}, lattice-based encryption~\cite{regev2006lattice}, and homomorphic signatures~\cite{lin2017linearly,gorbunov2015leveled,catalano2011adaptive}, NTT performance directly impacts end-to-end runtime. Enhancing NTT computation is thus critically important in the era of post-quantum cryptography and large-scale homomorphic computation.

Additionally, some cryptographic schemes require NTTs of multiple polynomial lengths simultaneously. For example, to exploit the computational advantages of both arithmetic FHE schemes (e.g., BFV \cite{fan2012somewhat}, BGV \cite{brakerski2014leveled}, and CKKS \cite{cheon2017homomorphic}) and logic FHE schemes (e.g., FHEW \cite{ducas2015fhew} and TFHE \cite{chillotti2020tfhe}) across different application scenarios, Hybrid Homomorphic Encryption (HHE) schemes \cite{lu2021pegasus,al2022openfhe,boura2020chimera} have emerged. However, designing an NTT hardware architecture for HHE is far from straightforward. In HHE, arithmetic FHE and logic FHE schemes require different NTT lengths. Designing separate NTT modules for these two types of FHE would result in prohibitive hardware overhead. Therefore, it is crucial to develop a unified NTT hardware architecture for HHE.

\begin{figure}[t]
    \centering
    \includegraphics[width=\linewidth]{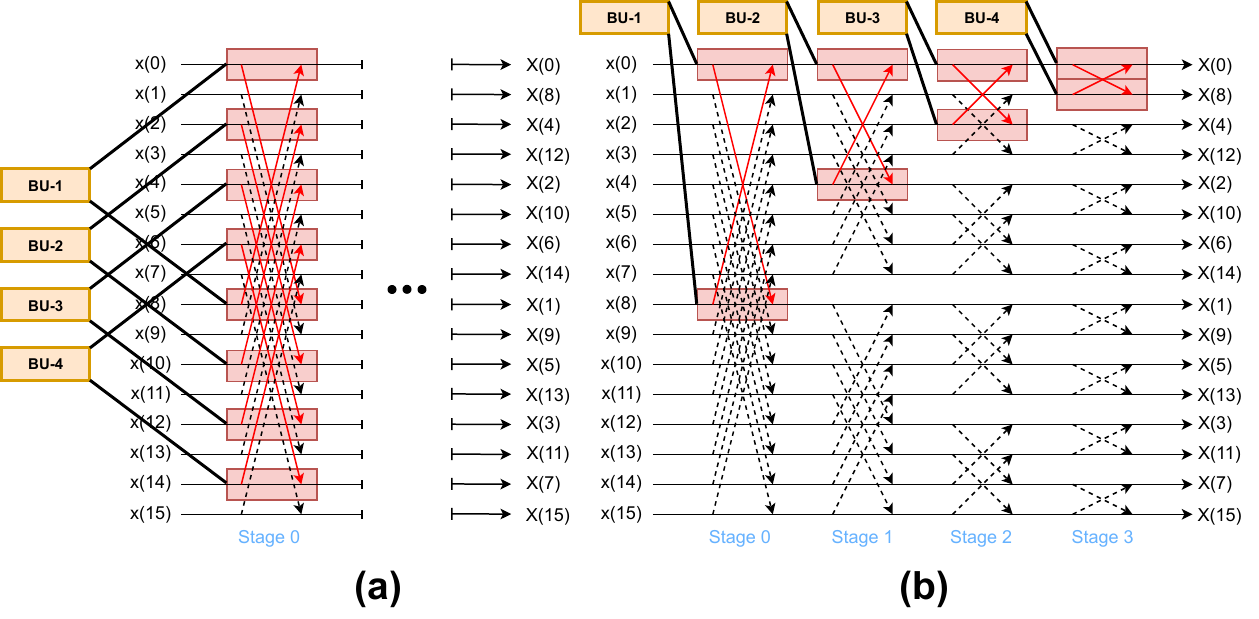}
    \vspace{-17pt}
    \caption{(a) Stage-based and (b) pipeline-based NTT architectures. Taking $N = 16$ and $p = 4$ as an example, the figure depicts the computational behavior of the $p$ butterfly units in the first iteration.}
\label{fig:two_NTT}
\vspace{-17pt}
\end{figure}

A considerable number of prior works have attempted to accelerate NTT computation across various hardware platforms \cite{zhang2024sok}, including GPU~\cite{jung2021over,fan2023tensorfhe}, FPGA~\cite{riazi2020heax,agrawal2023fab,yang2023poseidon,lu2024ntt}, and ASIC~\cite{kim2022ark,kim2022bts,samardzic2022craterlake,samardzic2021f1,deng2024trinity}. The hardware architectures for NTT computation are generally classified into two main categories: stage-based and pipeline-based designs. Figure \ref{fig:two_NTT} illustrates the data processed by $p$ parallel butterfly units during the first iteration for both architectures. In the stage-based NTT, multiple butterfly units are typically instantiated to operate in parallel, performing modular additions and multiplications on multiple data pairs within each transformation stage to reduce overall computation latency. This approach repeatedly reads and writes intermediate results to and from memory to achieve a trade-off between area and speed: a higher number of parallel butterfly units yields greater acceleration, but also increases on-chip data exchange and memory bandwidth requirements. At high levels of parallelism, ensuring timely data access while avoiding read-after-write (RAW) hazards becomes a significant challenge. On the other hand, the pipeline-based NTT connects butterfly units sequentially across computation stages, allowing data to flow through each stage in a pipeline manner, thereby reducing memory bandwidth demand. Because data flows sequentially, this type of architecture features a simple memory access pattern and avoids complex out-of-order accesses. However, its parallelism is limited and its structural flexibility constrained, making it difficult to fully exploit parallel potential. Existing studies have shown that pipeline-based NTT implementations can achieve high throughput, but often at the cost of significant hardware resource consumption, leaving room for further efficiency improvement.

In this paper, we present Hermes, a unified high-performance NTT architecture based on hybrid dataflow. Unlike stage-based and pipeline-based NTT architectures, where the former applies temporal parallelism to the stage dimension and spatial parallelism to the data dimension, and the latter applies them in the opposite way, Hermes applies both temporal and spatial parallelism to both dimensions, achieving superior throughput and scalability. A fully pipelined on-chip core guarantees high computational efficiency. For inherent NTT bank conflicts, our conflict-free on-chip fragmentation algorithm rearranges buffer data layout to enable burst HBM access. Lastly, an efficient dataflow works with this algorithm to increase computational intensity via data reuse, lowering HBM bandwidth needs. Our contributions can be summarized as follows:

\begin{itemize}
  \item We propose Hermes, a hybrid dataflow-based NTT architecture that performs parallel computation along both temporal and spatial dimensions.
  \item We implement efficient on-chip compute kernels with dedicated control circuitry, and we devise a conflict-free on-chip fragmentation algorithm to ensure maximal parallel read/write operations and eliminate pipeline bubbles.
  \item We propose an efficient dataflow that significantly enhances the compute intensity of on-chip kernels, thereby reducing HBM bandwidth consumption.
  \item We implement Hermes on a Xilinx Alveo U280 FPGA. Experimental results demonstrate high hardware utilization and reduced HBM bandwidth requirements, and Hermes achieves up to 13.6× and 1.3× higher throughput than state-of-the-art GPU and FPGA accelerators, respectively.
\end{itemize}

\section{Background and Related Work}
The Number Theoretic Transform (NTT), the finite-field analogue of the Discrete Fourier Transform (DFT) \cite{James1965CT,Gentleman1966GS}, is widely used to accelerate polynomial multiplication. In homomorphic encryption schemes, the NTT often becomes a major performance bottleneck due to its high arithmetic intensity and its large share of the total runtime. Consider two polynomials $\mathbf{a}, \mathbf{b} \in R_Q = \mathbb{Z}_Q[X]/(X^N + 1)$. A direct multiplication $\mathbf{c} = \mathbf{a} \cdot \mathbf{b}$ requires $\mathcal{O}(N^2)$ operations. By mapping the operands to the NTT domain, the convolution becomes element-wise multiplication, reducing the complexity to $\mathcal{O}(N \log N)$, i.e., $\mathrm{NTT}(\mathbf{a} \cdot \mathbf{b}) = \mathrm{NTT}(\mathbf{a}) \odot \mathrm{NTT}(\mathbf{b})$, where $\odot$ denotes element-wise multiplication. Hermes follows the optimized algorithm in Ref~\cite{longa2016speeding}, which uses a negative-wrapped convolution variant of the NTT to avoid zero padding. To enable such an NTT, we choose $N$ as a power of two and select a prime modulus $Q$ satisfying $Q \equiv 1 \pmod{2N}$, which guarantees the existence of a primitive $2N$-th root of unity $\psi \in \mathbb{Z}_Q$.  Moreover, since both the ciphertext modulus $Q$ and the twiddle-factor table $\psi$ are known in advance, we apply Shoup’s modular multiplication technique \cite{shoup2001ntl} to speed up the coefficient-twiddle products using a precomputed table $\psi_{\mathrm{rev}} \in \mathbb{Z}_Q^N$.

Due to the time-consuming nature of NTT in FHE schemes, previous works have focused on improving its throughput. Accelerators in \cite{jung2021over,fan2023tensorfhe} use GPUs, which offer high performance but are limited by a fixed architecture that hinders deep optimization for specific algorithms like NTT.

In contrast, FPGAs offer reconfigurability, enabling tailored data paths and pipelines for NTT optimization. \cite{riazi2020heax,yang2023poseidon,agrawal2023fab} present FPGA-based accelerators for arithmetic FHE. Among them,  HEAX \cite{riazi2020heax} adopts a general-purpose architecture with adjustable throughput, Poseidon \cite{yang2023poseidon} improves performance via employing an "NTT-fusion" technique, and FAB \cite{agrawal2023fab} deeply optimizes the NTT datapath. Ref \cite{lu2024ntt} proposes a state-of-the-art NTT-specific accelerator. Nonetheless, it exhibits poor data reuse and low computational intensity, with throughput constrained by HBM bandwidth. Furthermore, these designs do not support multiple polynomial lengths for NTT, which is unfavorable for HHE. Although HEAX \cite{riazi2020heax} supports automatic instantiation for different parameter sets, each configuration requires re-synthesizing and reprogramming the FPGA, making it unsuitable for dynamic parameter switching in practical HHE workloads. Recently, Ref \cite{deng2024trinity} introduced a design supporting multi-length NTT computation. However, it lacks an efficient dataflow design, leading to low throughput.

\section{Motivation}

\subsection{Roofline Model Analysis}
In this section, we analyze the performance bottlenecks of stage-based and pipeline-based NTT architectures. An NTT with a polynomial length of $N$ consists of $S = \log N$ stages. Each 2-point butterfly unit serves as the fundamental parallel computation element, and the architecture includes $p$ such units operating concurrently.

First, we analyze the theoretical upper bound of the computation intensity of the NTT. An $N$-point NTT requires loading at least $N$ input elements, giving a minimum memory traffic of $Q(N)=\mathcal{O}(N)$. The transform performs $N/2 \cdot \log N$ butterfly operations, each with constant computational cost, resulting in a total workload of $W(N)=\mathcal{O}(N\log N)$. According to the definition $I(N)=W(N)/Q(N)$, the theoretical upper bound of NTT computation intensity is therefore $\mathcal{O}(\log N)$.

In the stage-based architecture, each butterfly unit in any stage reads two input data elements, denoted $x_1$ and $x_2$, with $ b=2$. The unit size depends on the numerical precision. The additional memory access for twiddle factors is denoted as $bw$, where $w$ represents the number of twiddle factors involved, and this cost can be reduced to a constant through replication or recurrence relations. Hence, the total memory access volume for a complete $N$-point NTT computation is given by
$Q(N) = (2 + bw) \cdot p \cdot \frac{N}{p} \cdot \log N = \mathcal{O}(N \log N).$
It is straightforward to observe that the computational workload of the stage-based architecture is $W(N)=\mathcal{O}(N\log N)$. So the stage-based architecture exhibits a computational intensity of $\mathcal{O}(1)$.

In the pipeline-based architecture, $p$ must be an integer divisor of $\log N$. Otherwise, the $p$ butterfly units cannot be effectively pipelined. This constraint limits the scalability of the pipeline-based architecture, as blindly increasing parallelism quickly exhausts the available on-chip computing resources. Data passes through all $p$ stages of the pipeline, resulting in $2p$ data accesses. Similarly, the twiddle factor access is denoted as $bw$. Hence, the total memory access volume per pipeline pass is $Q(N) = 2 + bw = \mathcal{O}(1)$, while the computational workload per pipeline pass is $W(N) = \mathcal{O}(p) = \mathcal{O}(\log N).$ Therefore, the computational intensity of the pipeline-based architecture is $\mathcal{O}(\log N)$. Therefore, although the pipeline-based architecture achieves the theoretical maximum computation intensity, it is constrained by the number of NTT stages $S$, which limits its flexibility.

\begin{figure}[t]
    \centering
    \includegraphics[width=\linewidth]{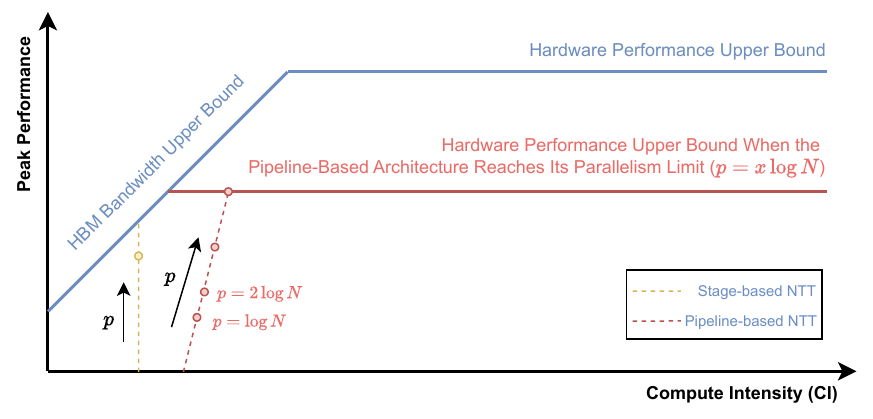}
    \vspace{-8pt}
    \caption{Roofline model of the two architectures.}
\label{fig:roofline}
\end{figure}

Their corresponding roofline models are shown in Figure \ref{fig:roofline}. The horizontal axis represents computational intensity, while the vertical axis indicates the achievable peak throughput. Furthermore, we illustrate how the performance of both architectures varies with increasing parallelism. For the stage-based architecture, the computational intensity remains constant, and as the number of 2-point butterfly units increases, the design quickly reaches the memory bandwidth limit. In contrast, for the pipeline-based architecture, the computation units are connected in a pipelined manner, allowing intermediate data to be reused on-chip. Consequently, as the parallelism increases in multiples of $S$, the computation intensity improves. However, the parallelism of the pipeline-based design is inherently constrained by the number of NTT stages $S$ and the available on-chip resources, making it difficult to configure the degree of parallelism flexibly.

\subsection{Design Opportunities}
From the design principles of the stage-based and pipeline-based architectures, it can be observed that:
(1) The stage-based architecture essentially organizes the computation by unfolding the stage and data dimensions of the NTT in the time domain, while unfolding the input data dimension in the spatial domain; (2) The pipeline-based architecture, in contrast, unfolds the input data dimension in the time domain and the stage dimension in the spatial domain, as summarized in Table \ref{tab:Design}.

\begin{table}[htbp]
\centering
\caption{Design Conceptual Differences between Different NTT Architectures}
\label{tab:Design}
\renewcommand{\arraystretch}{1.2}
\resizebox{0.48\textwidth}{!}{%
\begin{tabular}{c|cc|cc}
\toprule
\multirow{2}{*}{\textbf{Architecture Type}} &
\multicolumn{2}{c|}{\textbf{Stage Dimension}} &
\multicolumn{2}{c}{\textbf{Data Dimension}} \\
\cline{2-5}
& \textbf{Spatial} & \textbf{Temporal} & \textbf{Spatial} & \textbf{Temporal} \\
\midrule
Stage-Based & No & Yes & Yes & Optional \\
Pipeline-Based & Yes & No & Optional & Yes \\
Hybrid Dataflow-Based & Yes & Yes & Yes & Yes \\
\bottomrule
\end{tabular}
}
\end{table}

Both unfolding approaches have inherent drawbacks: the stage-based design not only suffers from low computational intensity but also experiences latency in individual NTT computations due to butterfly unit delays. Although the pipeline-based architecture eliminates butterfly unit latency by achieving an iteration interval of one, its performance is constrained by the number of NTT stages $S$ and the
available on-chip resources.

\begin{figure}[t]
    \centering
    \includegraphics[width=0.9\linewidth]{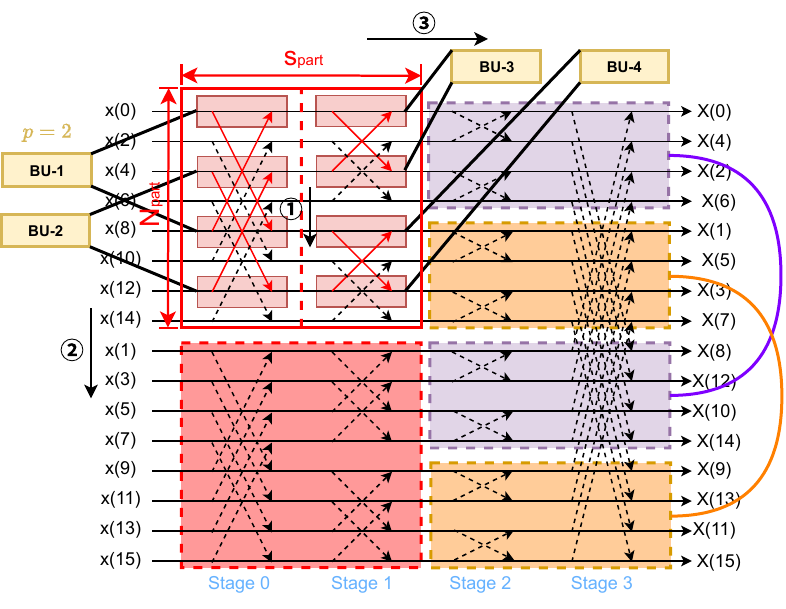}
    \vspace{-8pt}
    \caption{Hybrid dataflow-based NTT architecture. Here, we set \(N_{part}=4, S_{part}=2, p=2\). The input and output orders are rearranged for illustration.}
\label{fig:hybrid NTT}
\end{figure}

Therefore, we propose an NTT architecture based on hybrid dataflow, as illustrated in Figure \ref{fig:hybrid NTT}. In this design, both key dimensions of NTT computation, the data dimension (path \textcircled{2}) and the stage dimension (path \textcircled{3}), are unfolded simultaneously in both the temporal and spatial domains. Parallelism along the data dimension processes $N_{\mathrm{part}}$ points, where these $N_{\mathrm{part}}$ points span $S_{\mathrm{part}}$ partial stages along the stage dimension, with each sub-stage handled in parallel by $p$ butterfly units. Although the computational intensity of the hybrid dataflow architecture remains theoretically at the same order as that of the pipeline-based design, i.e., $\mathcal{O}(\log p)$, the introduction of an additional parallel dimension removes the dependency of the parallel unit limit on the number of stages.

Meanwhile, the processing order of the instantiated NTT engines (highlighted in the big box) is constrained, where the input data within each big box are packed into the corresponding pipeline positions of the butterfly units, enabling efficient computation through a block-level parallel pipelining scheme. As a result, the design achieves an iteration interval (II) of one, similar to that of the pipeline-based architecture. Specifically, each data block encapsulates $s$ data elements and is processed in a task-level pipeline manner along path \textcircled{1}. Although the task iteration interval (taskII) equals $s$, the overall throughput satisfies $s/\text{taskII} = 1$, thereby maintaining full computational efficiency.Overall, the hybrid dataflow operates in the order of paths \textcircled{1}, \textcircled{2}, and \textcircled{3}.

However, in cases where the number of stages $S$ is not divisible by the degree of spatial parallelism, the hybrid dataflow architecture requires additional control units, which will be discussed in Section~\ref{subsec:architecture}. In addition, because the hybrid dataflow architecture performs both spatial and temporal unfolding along the stage and data dimensions, the heterogeneous stage index strides during parallel computation lead to memory bank conflicts, which will be addressed in Section~\ref{subsec:algorithm}.

\section{Hermes Methodology}
\begin{figure}[t]
    \centering
    \includegraphics[width=\linewidth]{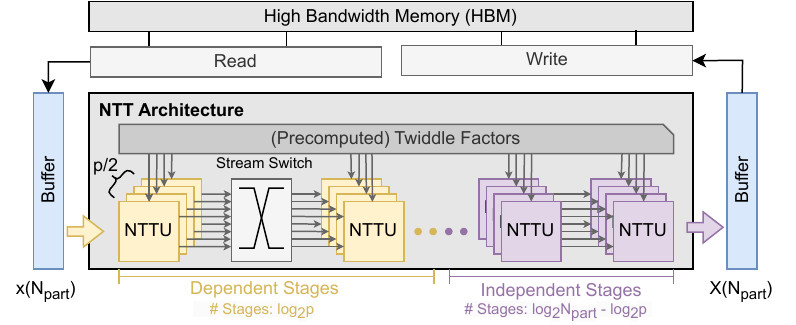}
    \caption{The overall architecture of Hermes including dependent and independent stages with $p/2$ NTT units per stage.}
\label{fig:NTT-Architecture}
\end{figure}

Hermes is a unified NTT architecture with hybrid dataflow supporting multiple polynomial lengths via configurable computation modes, enabling varied-length NTT without notable hardware overhead. Its conflict-free on-chip fragmentation algorithm, integrated with optimized dataflow, balances computation and memory-access latencies, thereby achieving high throughput.

\subsection{Proposed Architecture}
\label{subsec:architecture}

We propose a unified NTT architecture with hybrid dataflow supporting variable-length polynomials for the CKKS and TFHE schemes. Figure \ref{fig:NTT-Architecture} illustrates the complete NTT computation process. Each iteration computes a segment of the polynomial vector of length $N$, denoted $x(N_{part})$, spanning $S_{part} = \log(N_{part})$ stages. With $p/2$ NTT Units per stage, a total of $p/2 \times \log(N_{part})$ units are instantiated on-chip. Precomputed twiddle factors are loaded from HBM into BRAM, supplying data to NTT and butterfly units. URAM acts as scratchpad memory for intermediate polynomial coefficients.

Figure \ref{fig:NTT-internal} details the internal structure of the NTT Unit (NTTU hereafter) and Butterfly Unit (BU hereafter).

\textbf{\textit{Butterfly Unit}}. The BU is Hermes' minimal computational block, executing a fully pipelined butterfly operation enhanced by Shoup's Technique \cite{shoup2001ntl} for modular multiplication. The BU operates in two configurable modes: (1) \textit{Butterfly Mode} for standard operations, and (2) \textit{Swap Mode} for bypassing the butterfly, directly swapping coefficient pairs.

\textbf{\textit{NTT Unit}}. The NTTU, serving as the fundamental computation module in Hermes, comprises a Control Unit, a Data Read Unit, and two BUs. The Control Unit sets the operational mode, indicating how many final stages employ \textit{Butterfly Mode}. The Data Read Unit manages polynomial coefficients and twiddle factors based on the NTTU's stage and index. After BU processing, resulting data streams advance to the subsequent NTTU.

\subsection{Efficient Dataflow}
\begin{figure}[t]
\vspace{-12pt}
    \centering
    \includegraphics[width=\linewidth]{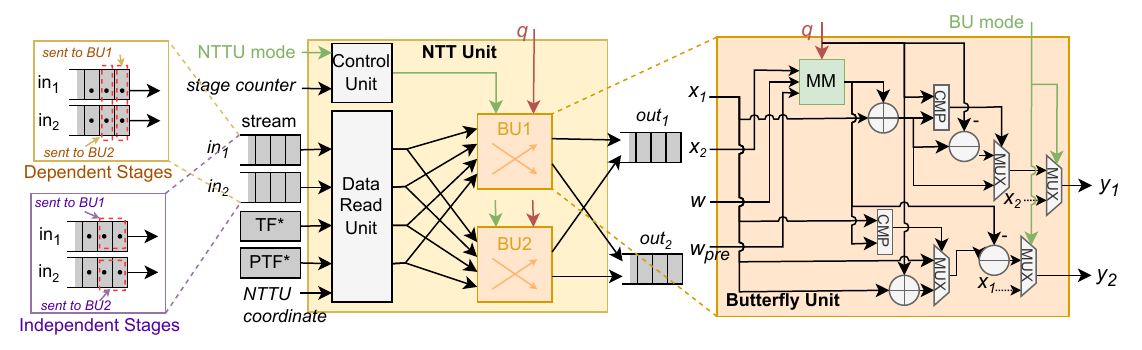}
    \vspace{-15pt}
    \caption{Internal computation units of NTT (NTTU, BU).}
\label{fig:NTT-internal}
\end{figure}

The NTT architecture design leverages the index difference in the butterfly operation, \(\frac{N_{\mathrm{part}}}{2^{i+1}}\), which decreases as stages progress, transitioning from \textit{independent} to \textit{dependent} stages, as shown in Figure \ref{fig:NTT-Architecture}. In each stage with \(p/2\) NTTUs, the number of dependent stages is \(\log p \). In dependent stages, two input streams are assigned to BU1 and BU2, while in independent stages, \({in}_1\) and \({in}_2\) are mapped to BU1 and BU2, respectively, as shown in Figure \ref{fig:NTT-internal}.

\begin{figure}[t]
    \centering
    \includegraphics[width=\linewidth]{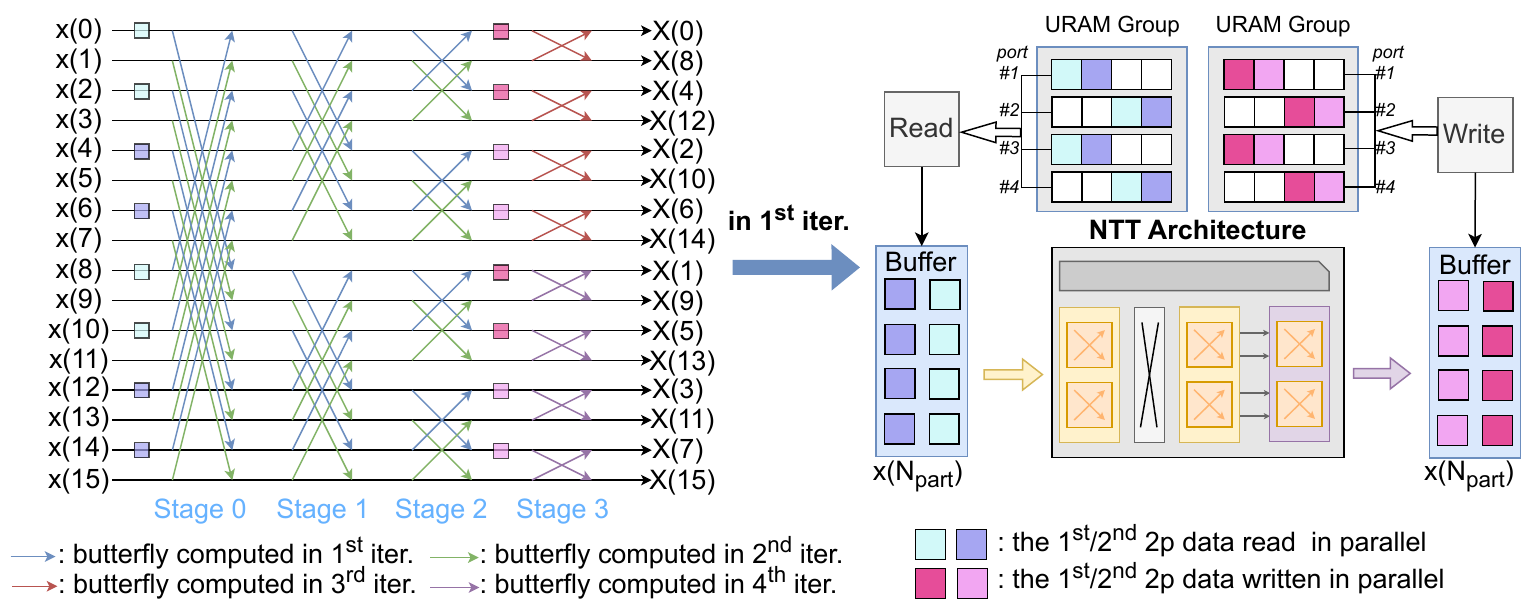}
    \vspace{-20pt}
    \caption{An example of iteratively computing an \(N\)-point NTT using an \(N_{part}\)-point NTT with $p/2$ NTTUs in each stage. Here, we set $p$ to 2, \(N_{part}\) to 8 and \(N\) to 16, resulting in 4 iterations. In Hermes, $p$ is set to 16 and \(N_{part}\) is set to 256.}
    \label{fig:DAP2}
\vspace{-5pt}
\end{figure}

For larger transforms (\(N > N_{\mathrm{part}}\)), \(\frac{2N}{N_{\mathrm{part}}}\) iterations of an \(N_{\mathrm{part}}\)-point NTT are performed. The left side of Figure \ref{fig:DAP2} illustrates a schematic of iteratively computing a 16-point NTT using an 8-point NTT with 1 NTTU in each stage. (i.e., \(N_{\mathrm{part}} = 8\), \(N = 16\), \(p = 2\)). In the first half, BUs remain in \textit{Butterfly Mode}, while in the second half, only the final stage uses \textit{Butterfly Mode}, with \textit{Swap Mode} enabled otherwise. Formally, if \(S = \log N\), \textit{Butterfly Mode} is enabled for the last \(S - S_{\mathrm{part}} \cdot \lfloor (S-1)/S_{\mathrm{part}} \rfloor\) stages. 

The right side of Figure \ref{fig:DAP2} presents the detailed dataflow of the first iteration when one NTTU is instantiated in each stage. For each cycle, Hermes reads $2p$ data elements from the URAM in parallel and feeds them into the NTTUs for computation. The on-chip buffer data layout is defined by the Conflict-Free On-Chip Fragmentation Algorithm (see Section \ref{subsec:algorithm} for details), and after passing through \(S_{part}\) stages, results are written back to the URAM with a parallelism factor of $2p$. The initiation interval for URAM read/write operations matches that of the NTTUs, guaranteeing efficient dataflow and preventing pipeline stalls. Upon completion of the first iteration, the results are retained in the on-chip URAM to serve as inputs for the third and fourth iterations. By caching intermediate results in the on-chip URAM, we improve data reuse and computational intensity, reducing Hermes’s demand on HBM bandwidth.

\subsection{Conflict-Free On-Chip Fragmentation Algorithm}
\label{subsec:algorithm}

\begin{figure}[t]
\vspace{-12pt}
    \centering
    \includegraphics[width=\linewidth]{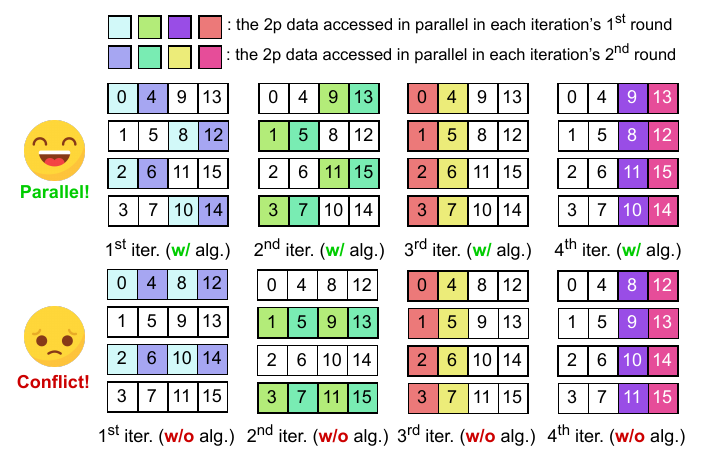}
    \vspace{-10pt}
    \caption{URAM data arrangement and the parallel read/write patterns over 4 iterations (w/ and w/o mapping algorithm), for \(N_{\mathrm{part}} = 8\), \(N = 16\), \(p = 2\).}
\label{fig:data_arrangement}
\vspace{-12pt}
\end{figure}

In the NTT computation process, the spacing between the two data elements required by each butterfly operation (i.e., the stage stride) varies across stages. As a result, in the first half of the iterations, data accesses must satisfy a pairwise spacing of $N/2$, whereas in the second half, data will be accessed sequentially (in Hermes, since $N_{\mathrm{part}} > 2^{\frac{S-1}{2}}$, the stage stride in the latter iterations is smaller than $N_{\mathrm{part}}$). To fully exploit HBM bandwidth, burst read and write operations must be initiated sequentially. Moreover, NTT exhibits strict data dependencies: the input of each stage is the output of the preceding one. To avoid pipeline stalls due to memory access latency, we perform parallel read/write operations of $2p$ data elements per cycle to satisfy the requirement of $p/2$ NTTUs per stage.

In summary, we must design a conflict-free on-chip fragmentation algorithm that arranges data in the on-chip buffer to simultaneously satisfy the following conditions: (1) support both access patterns of the two kinds of iterations; (2) enable efficient data transfer between the parallel on-chip buffer and HBM in burst mode; (3) enable parallel read/write of $2p$ data elements. The algorithm used by Hermes is illustrated in Algorithm \ref{alg:mapping}. Its inputs are Hermes’s three configuration parameters, $N$, $N_{\mathrm{part}}$, and $p$. And its output is a two-dimensional array representing the buffer layout of the $N$ data elements, which satisfies the aforementioned three requirements.
\vspace{-6pt}
\begin{algorithm}
    \caption{Conflict-Free On-Chip Fragmentation Mapping}
    \label{alg:mapping}
    \begin{algorithmic}
        \REQUIRE {$N$, $N_{\mathrm{part}}$, $p$}
        \ENSURE $\mathit{data\_arrangement}[2p][N/(2p)]$
        
\FOR{$i \leftarrow 0$ \TO $N-1$}
    \STATE $bank\_id \leftarrow \bigl(i \oplus \lfloor i/N_{\mathrm{part}}\rfloor\bigr)\bmod (2p)$
    \STATE $offset \leftarrow i/(2p)$
    \STATE $\mathit{data\_arrangement}[bank\_id][offset] \leftarrow i$
\ENDFOR
    \RETURN $\mathit{data\_arrangement}$
    \end{algorithmic}
\end{algorithm}
\vspace{-6pt}

Hermes must support NTT computations of up to $2^{16}$ length, with $p$ set to 16. This implies that each parallel buffer must store at most $2^{16}/(2p)=2048$ elements, which fits entirely within a Xilinx U280 URAM. Hence, we choose URAM for data storage. Figure \ref{fig:data_arrangement} presents an instance of this algorithm’s application. It shows the data arrangement within the parallel URAMs and the parallel read/write patterns across $\tfrac{2N}{N_{\mathrm{part}}}$ iterations. Each iteration comprises $\frac{N_{\mathrm{part}}}{2p}$ data read/write rounds, distinguished by elements in distinct colors.

\subsection{Twiddle Factors Arrangement}

\begin{figure}[t]
\vspace{-5pt}
    \centering
    \includegraphics[width=\linewidth]{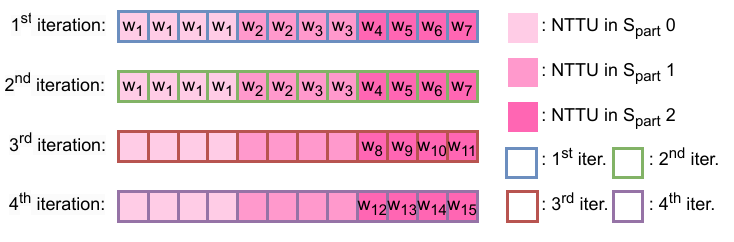}
    \caption{The arrangement of twiddle factors in 4 iterations across 3 partial stages when $N_{part} = 8$ and $ N=16 $. The empty boxes indicate that the BU mode in the NTTU is configured to the \textit{Swap Mode}.}
    \label{fig:TF}
\vspace{-5pt}
\end{figure}

An \(N_{\mathrm{part}}\)-point NTT requires \(N_{\mathrm{part}} - 1\) distinct twiddle factors, which for large \(N\) must reside in off-chip HBM and be loaded into on-chip BRAM at the start of computation. In each stage, the \(p/2\) NTTUs may demand the same twiddle factor, storing only one copy per distinct factor would bottleneck throughput. To avoid this, we replicate twiddle factors, resulting in a total stored copy count of only \(\tfrac{\log(N_{\mathrm{part}})}{2}\) times the original count, which ensures each NTTU in every partial stage has an independent copy and preserves parallelism. Figure~\ref{fig:TF} shows the twiddle factor arrangement within each NTTU for all iterations, and the precomputed twiddle factors follow the same arrangement.

\section{Experimental Results}

 \subsection{Experimental Setup}
 \label{subsec:setup}
 Our experimental platform is a server equipped with a Xilinx U280 FPGA and an Intel(R) Xeon(R) Platinum 8272CL CPU @ 2.60GHz. The FPGA chip comprises 1,304k LUTs, 2,607k FFs, 4,032 BRAMs, 960 URAMs, and 9,024 DSPs. We implemented Hermes on the U280 using the Xilinx Vitis 2023.1 toolchain, operating at a frequency of 300 MHz. In Hermes, we set \(N_{part}\) to 256, and \(p\) to 16. The polynomial coefficients in Hermes have a bit-width of 64 bits. We conduct a hardware utilization experiment across nine polynomial lengths ($N=2^8$ through $2^{16}$) and a performance experiment on the latter four ($N=2^{13},2^{14},2^{15},2^{16}$). According to the HE Security Standard White Paper \cite{albrecht2021homomorphic}, all parameter sets we use achieve a security level of 128-bit. The BU modes follow a consistent configuration pattern. In each iteration, the partial stages are divided into $i$ front sub-stages operating in \textit{Swap Mode} and $j$ latter sub-stages operating in \textit{Butterfly Mode}, where $i + j = S_{\text{part}}$. This configuration is denoted as “$S \times i, B \times j$”. For example, for the $2^{13}$-point NTT, the configuration is “$S \times 0, B \times 8$” during the first half of the iterations, and “$S \times 3, B \times 5$” during the second half. We generate twiddle factors and their precomputed values via OpenFHE~\cite{al2022openfhe}. Our resource utilization is shown in Table \ref{tab:resource}.

\begin{table}[t]
\centering
\caption{FPGA resource usage of Hermes.}
\vspace{-10pt}
\label{tab:resource}
\begin{tabular}{c c c c c c}
\toprule
\textbf{Component} & \textbf{BRAM} & \textbf{DSP} & \textbf{FF (k)} & \textbf{LUT (k)} & \textbf{URAM} \\
\midrule
\textbf{Hermes} & 68 & 6144 & 1{,}143 & 738 & 128 \\
---NTTU  & 0  & 96   & 9.8   & 4.7   & 0 \\
---BU   & 0  & 48   & 2.4  & 1.7 & 0 \\
\bottomrule
\end{tabular}
\end{table}


\subsection{Hardware Utilization}

To validate the performance advantages of the Hermes architecture across NTT computations of varying lengths, we evaluated its resource utilization for polynomial sizes from $2^8$ to $2^{16}$ and compared the results with those of the FPGA accelerator FAB and the ASIC accelerator Trinity \cite{deng2024trinity}, as shown in Figure \ref{fig:utilization}. The experimental results are as follows: (1) For all polynomial sizes, Hermes achieves significantly higher utilization than FAB. (2) When $N$ is small, Trinity performs best due to its highly flexible computational architecture. However, this advantage comes at the cost of significantly greater hardware overhead. (3) As $N$ increases, the utilizations of Hermes and Trinity become nearly identical. Because FHE schemes typically require the polynomial degree $N$ to be at least $2^{15}$ for security, Hermes’s resource utilization is sufficiently high for practical applications.

\begingroup
\setlength{\abovedisplayskip}{-2pt}
\setlength{\belowdisplayskip}{-20pt}

\begin{figure}[t]
\vspace{-6pt}
    \centering
    \includegraphics[width=\linewidth]{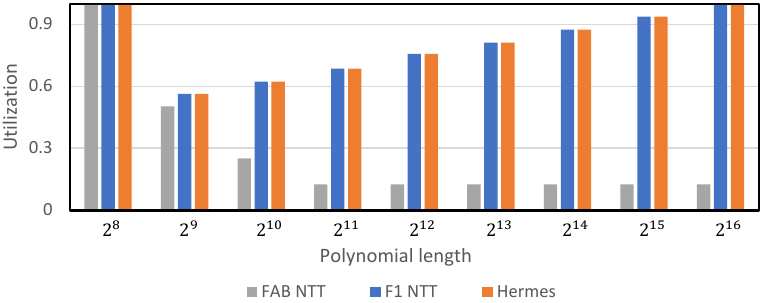}
    \vspace{-20pt}
    \caption{Hardware utilization.}
\label{fig:utilization}
\vspace{-10pt}
\end{figure}

\endgroup

\begin{figure}[t]
    \centering
    \includegraphics[width=\linewidth]{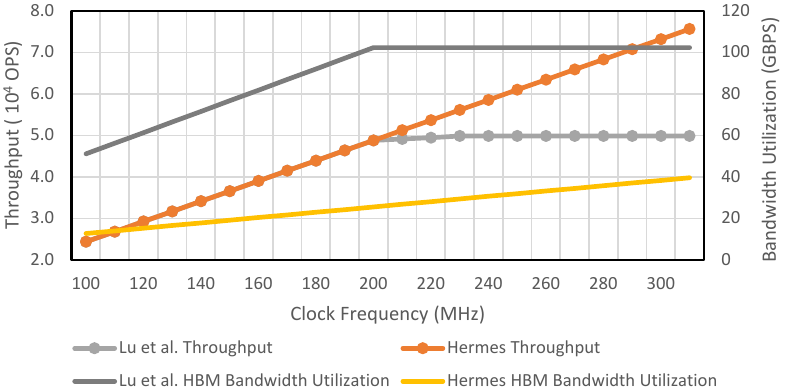}
    \vspace{-10pt}
    \caption{Bandwidth utilization and throughput comparison between Hermes and Ref~\cite{lu2024ntt}.}
\label{fig:utilization2}
\vspace{-12pt}
\end{figure}

\subsection{Performance}
Table \ref{tab:NTT performance} presents a throughput comparison between the proposed NTT architecture and previous schemes. Since a polynomial length of \(N=2^{16}\) meets the security and computational requirements of almost all real-world applications, we perform the comparison with previous schemes using this parameter. For schemes whose performance is not reported under certain NTT lengths, the corresponding data are obtained from Refs \cite{yang2023poseidon, lu2024ntt}. Additionally, because the proposed architecture supports NTT computations for different polynomial sizes, we also evaluate four distinct values of \(N\), as shown in Table~\ref{tab:NTT performance}.


While CUDA offers strong general-purpose performance and ease of use, the fixed GPU architecture cannot be extensively optimized for specific algorithms such as the NTT, thereby limiting its acceleration efficiency. In contrast, FPGAs are reconfigurable, allowing datapaths and pipelines to be tailored to algorithmic needs. Hermes achieves 17.7$\times$ and 13.6$\times$ higher throughput compared to the GPU-based designs in Ref \cite{jung2021over,fan2023tensorfhe} using V100 and A100, respectively.

Hermes achieves a 1.3× improvement in NTT performance compared to state-of-the-art scheme \cite{lu2024ntt}. Due to the high degree of parallelism in our architecture, it utilizes more hardware resources (especially DSPs) compared to other schemes, but achieves significant performance improvements. We have designed a highly efficient dataflow to balance the computation and memory access loads. Moreover, the internal computation units are fully pipelined.

We also compare the performance across multiple parameter sets. As described in Section \ref{subsec:setup}, from $N={2^{16}}$ to $N={2^{13}}$, the number of stages in which the BU is configured to \textit{Butterfly Mode} decreases in the latter half of the iterations, requiring additional time for data exchanges in \textit{Swap Mode}. The experimental results indicate that as \(N\) decreases, the throughput improvement factor also diminishes. Such a performance loss is entirely acceptable, as our unified hardware architecture supports NTT of varying lengths without requiring additional circuitry, thereby achieving higher area efficiency.

\begin{table}[t]
\setlength{\tabcolsep}{5pt} 
\centering
\vspace{-10pt}
\caption{Comparisons of NTT performance with previous schemes.}
\vspace{-6pt}
\label{tab:NTT performance}
\resizebox{1.0\linewidth}{!}{
\begin{tabular}{c c c c c c c}
\toprule
\textbf{Scheme} & \textbf{Platform} & \makecell{\textbf{Word}\\\textbf{Length}} & \makecell{\textbf{Freq.}\\\textbf{(MHz)}} & \textbf{DSP} & \textbf{N} & \makecell{\textbf{Throughput}\\\textbf{(OPS)}} \\
\midrule
100x~\cite{jung2021over}         & V100       & 52-bit & 1,245 & / & $2^{16}$ & 3,619  (17.7$\times$)\\
TensorFHE~\cite{fan2023tensorfhe}& A100       & 32-bit & 1,065 & / & $2^{16}$ & \textbf{4,705  (13.6$\times$)}\\

\midrule
\multirow{3}{*}{HEAX \cite{riazi2020heax}} 
 & \multirow{3}{*}{Stratix10} & \multirow{3}{*}{54-bit} & 300 & / & $2^{16}$ & 237 (270.8$\times$)\\
 &                            &                      & 300 & 2370 &$2^{14}$ & 41,853 (4.5$\times$) \\
 &                            &                      & 300 & 2610 &$2^{13}$ & 90,144 (3.1$\times$)\\

\midrule
FAB~\cite{agrawal2023fab}        & U280       & 54-bit & 300  & / & \textbf{$2^{14}$} & 167,000 (1.1$\times$)    \\
Poseidon~\cite{yang2023poseidon} & U280       & 32-bit & 450  & 4032 & $2^{16}$ & 12,474 (5.1$\times$)\\
Lu et al.~\cite{lu2024ntt}       & U55C       & / & 200  & / & \textbf{$2^{16}$} & \textbf{48,543 (1.3$\times$)} \\

\midrule
\textbf{\multirow{4}{*}{Ours}} 
 & \multirow{4}{*}{U280} & \multirow{4}{*}{64-bit} & 300 & \multirow{4}{*}{6144} & \textbf{$2^{16}$} & \textbf{64,172} \\
 &                       &                      & 300 &                      & $2^{15}$ & 114,330 \\
 &                       &                      & 300 &                      & $2^{14}$ & 187,500 \\
 &                       &                      & 300 &                      & $2^{13}$ & 275,736 \\
\bottomrule
\end{tabular}
}
\vspace{-10pt}
\end{table}


\subsection{Relationship Between Bandwidth Utilization and Performance}
We evaluated the peak throughput and bandwidth utilization of Hermes across operating frequencies ranging from 100 MHz to 300 MHz, and compared the results with a state-of-the-art FPGA-based NTT accelerator \cite{lu2024ntt}, as shown in Fig.~\ref{fig:utilization2}. When the clock frequency exceeds 200 MHz, the throughput of Ref~\cite{lu2024ntt} reaches its maximum value (48,543 OPS) and ceases to increase further, due to the inherent latency associated with HBM access. The bandwidth utilization of this design also peaks at 200 MHz.

In contrast, Hermes adopts a hybrid dataflow architecture, in which data fetched from HBM are stored in scratchpad memory for subsequent reuse by the computation units. Hermes requires significantly less bandwidth than Ref~\cite{lu2024ntt}, which directly contributes to the superior efficiency of the proposed architecture. Due to pipeline filling and draining overheads, the actual performance of Hermes (Table~\ref{tab:NTT performance}) is lower than the theoretical peak values reported here.

\section{Conclusion}
In this paper, we present Hermes, a unified NTT architecture with hybrid dataflow supporting various polynomial lengths for HHE. We design a conflict-free on-chip fragmentation algorithm for parallel data access, and an efficient dataflow that balances computation and memory access to maximize the throughput. Future work will focus on identifying the optimal trade-off between resource usage and throughput, as well as supporting additional computational primitives for the CKKS and TFHE schemes to accelerate HHE applications.

\bibliographystyle{ACM-Reference-Format}
\bibliography{ref}

\appendix









\end{document}